\documentclass[12pt]{iopart}
\usepackage{graphicx,amssymb,ulem}
\usepackage[T1]{fontenc}
\usepackage[english]{babel}
\usepackage{float}
\usepackage{xcolor}
 \expandafter\let\csname equation*\endcsname\relax
\expandafter\let\csname endequation*\endcsname\relax
\usepackage{amsmath}

\usepackage{url}

\DeclareMathOperator{\sinc}{sinc}

\begin{document}

\title{Spatial entanglement and state engineering via four-photon Hong-Ou-Mandel interference}
\author{A.~Ferreri$^1$, V.~Ansari$^1$, B.~Brecht$^1$, C.~Silberhorn$^1$, P.~R.~Sharapova$^1$}
\address{$^1$ Department of Physics and CeOPP, University of Paderborn, Warburger Strasse 100, D-33098 Paderborn, Germany}

\begin{abstract}
The phenomenon of entanglement is the basis of quantum information and quantum communication processes. Entangled systems with a large number of photons are of great interest at present because they provide a platform for streaming technologies based on photonics.  In this paper we present a device which operates with four-photons and based on the  Hong-Ou-Mandel (HOM) interference.  The presented device  allows to maximize the degree of spatial entanglement and generate the highly entangled four-dimensional Bell states. Furthermore, the use of the interferometer  in different regimes leads to fast interference fringes in the coincidence probability with period of oscillations twice smaller than the pump wavelength. We have a good agreement between theoretical simulations and experimental results.
\end{abstract}
\maketitle

\section{Introduction}

Entanglement is a fundamental element of quantum physics \cite{schrodinger1935gegenwartige} which is widely used in quantum optical technologies, such as quantum cryptography \cite{Grblacher2006, PhysRevLett.67.661}, quantum algorithms \cite{PhysRevLett.111.100501, PhysRevA.67.042304} quantum computing \cite{DiVincenzo255, Makhlin2002} and quantum teleportation \cite{bouwmeester1997experimental, Furusawa706}.
Entanglement can be created by using non-linear photon sources such as parametric down-conversion (PDC) process \cite{rubin1994theory, law2004analysis, jeronimo2009control} or by  using different optical elements \cite{PhysRevA.96.043861, PhysRevA.75.034309, PhysRevA.66.014102}, such that the output photons can be related by frequency,  polarization \cite{PhysRevLett.92.153602} or spatial   \cite{simon2002polarization, li2010deterministic} entanglement.


A typical tool for creating a spatial entanglement is the well-known two-photon Hong-Ou-Mandel (HOM) interference \cite{PhysRevA.96.043857, hong1987measurement, PhysRevA.87.062106, kim2016two, tichy2014interference}, which generates high-entangled NOON states with N=2. The NOON states, a class of maximized entangled states, are extremely important in quantum lithography \cite{PhysRevLett.85.2733} and quantum metrology, due to the possibility of overcoming the standard quantum limit \cite{doi:10.1080/00107510802091298}. Although the multiphoton HOM interference cannot deterministically generate NOON states with N>2,   it could be the best candidate for creating other high-entangled photon states, such as higher-dimensional Bell states \cite{JAEGER2004425, Jaeger2008, Wang:17, PhysRevA.71.032308} and Greenberger-Horne-Zeilinger (GHZ) states \cite{bouwmeester1999observation, pan2000experimental, hillery1999quantum}.

The multiphoton interference and generation of high-entangled multidimensional states are of great interest at present \cite{Bell:19, metcalf2013multiphoton} because of their applications in various areas of quantum optics,  such as boson sampling \cite{wangh}, quantum walks \cite{Gard:13}, quantum metrology \cite{Nagata726}, quantum communications  \cite{PhysRevA.86.010302}. Recently, a new strategy based on the spectral engineering of the photon sources seems to open new frontiers to manipulate and control the interference process, for example by exploiting the frequency correlations of the generated entangled photons \cite{PhysRevLett.121.033601}, or even by manipulating the temporal modes structure of the source  \cite{PhysRevA.100.053829}. 

In this paper we investigate spatial entanglement generated in the four-photon HOM interferometer with a polarization rotation. The proper conditions to create maximally entangled states were found. The coincidence probabilities to detect different amount of photons in different channels are calculated and measured. The antibunching behaviour as well as the fast oscillations in the coincidence probability are observed and analysed.  Generation of various four-dimensional Bell states combinations at different path delays is shown. Creation of fast oscillations with a period twice smaller than the pump wavelength inside the interferometer is presented.


\section{Theoretical model}

\begin{figure}[H]
\includegraphics[width=1\textwidth]{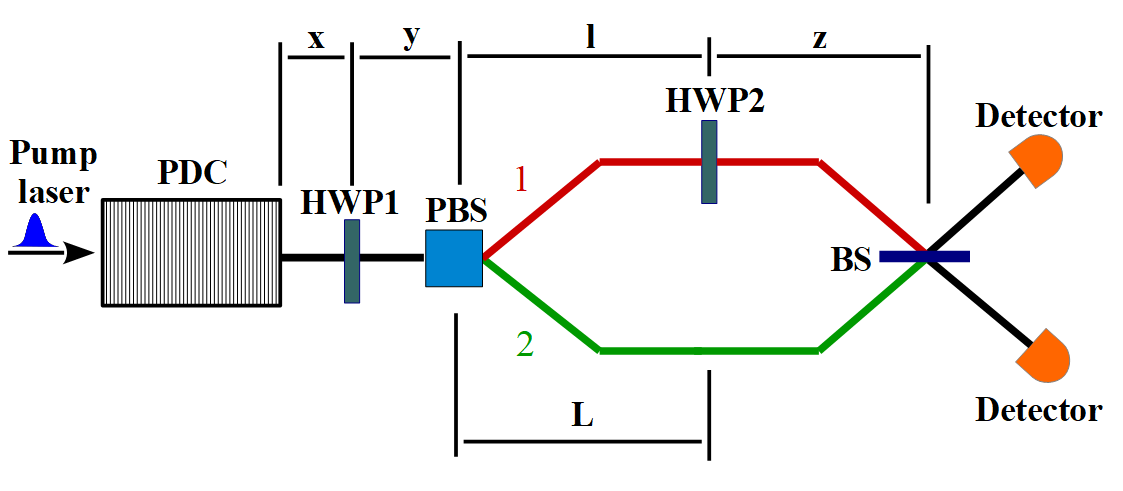}
\caption{Schematic setup. Four photons are generated in the PDC section.  For every photon, a first half-wave plate, $HWP_1$ creates  a superposition of the vertical and the horizontal polarizations. A polarization beam slitter routs vertically and horizontally polarized photons in two different spatial channels (red and green). The second half-wave plate $HWP_2$  rotates the polarization in the first channel, such that all photons own the same polarization. A path length of the second channel compensates arising time delay between the vertically and horizontally polarized photons $L=l+\Delta l$. Finally, all photons interact in a beam splitter and are detected.}
\label{phim}
\end{figure}

In this work we consider a linear quantum interferometer with a polarization roatation \cite{PhysRevA.96.043857} in the four-photon  regime. The schematic setup is presented in  Fig.\ref{phim}, and includes the four-photon source (type-II PDC process in the periodic poled Potassium titanyl phosphate material, ppKTP), the interferometric part and the detection stage. 

In the low intensity regime,  the four-photon state generated in the type-II PDC process can be calculated by using the second order perturbation theory without taking into account the time ordering effect \cite{PhysRevA.100.053829} (see also Appendix A). The half-wave plate $HWP_1$ with the conversion parameter $\phi_1$ is placed after the four-photon source and creates a superposition of the vertically and horizontally polarizations of both the signal and idler photons. Then, the polarization-sensitive beam splitter ($PBS$) splits the state having different polarizations in the two spatially separated channels. Afterwards, the second half-wave plate $HWP_2$ switches perfectly the polarization in the first channel. The path delay is implemented in the second channel. Finally, the quantum interference occurs in the beam splitter ($BS$) with the rotation angle $\theta$, and the coincidence probability is measured by the detection stage.
The all mentioned steps can be mathematically described by corresponding unitary transformation matrices related to the elements inside the interferometer:
\begin{equation}
U = BS\cdot FP_3\cdot HWP_2\cdot FP_2\cdot PBS\cdot FP_1\cdot HWP_1\cdot FP_0,
\label{utot}
\end{equation}
where matrices  $FP_x$ describe the free propagation of the photons in the air between various physical elements \cite{sharapova2017toolbox}. 

To make an entanglement analysis more evident, it is better to start from the wave function before the last beam splitter $|\psi_{before BS}\rangle$. This wave function can be obtained by acting the unitary transformation Eq.\ref{utot} without the $BS$ matrix on the input state, see Appendix A. 
Performing the Schmidt decomposition of the joint spectral amplitude (JSA), one can introduce two sets of the broadband modes: the original ones:
\begin{equation}\begin{split}
A^\dagger_k=\int d\omega_s u_k(\omega_s) a^\dagger(\omega_s) \\
B^\dagger_k=\int d\omega_i v_k(\omega_i) a^\dagger(\omega_i).
\end{split}
\end{equation} 
and the path-dependent modes:
\begin{equation}\begin{split}
C^\dagger_k=\int d\omega_s u_k(\omega_s)e^{i \omega_s\frac{\Delta l}{c}}   a^\dagger(\omega_s) ,\\
D^\dagger_k=\int d\omega_i v_k(\omega_i) e^{i \omega_i\frac{\Delta l}{c}}  a^\dagger(\omega_i),
\end{split}
\label{Schmidt_CD}
\end{equation} 
where $\Delta l$ is the path delay in the second arm of the interferometer, $c$ is the speed of light, $u_k(\omega_s)$ and $v_k(\omega_i)$ are the eigenfunctions of the Schmidt decomposition, see Appendix A.

The wavefunction before the beam splitter, see Eq. (\ref{part11}-\ref{part33}), can be written in terms of the broadband Schmidt modes:

\begin{equation}
|\psi_{before BS}\rangle=|\psi_{before BS}(22)\rangle \\+|\psi_{before BS}(4004)\rangle+|\psi_{before BS}(3113)\rangle,
\label{total}
\end{equation}
with

\begin{equation}
\begin{split}
|\psi_{before BS}(22)\rangle= \sum_{k \tilde k}\sqrt{\lambda_k\lambda_{\tilde k}} 
\bigg[-C_{k_2}^\dagger B_{k_1}^\dagger C_{\tilde k_2}^\dagger  B_{\tilde k_1}^\dagger\sin^4\phi_1 +C_{k_2}^\dagger B_{k_1}^\dagger  A_{\tilde k_1}^\dagger  D_{\tilde k_2}^\dagger \sin^2\phi_1\cos^2\phi_1 \\
+A_{k_1}^\dagger  B_{k_1}^\dagger C_{\tilde k_2}^\dagger  D_{\tilde k_2}^\dagger\sin^2\phi_1\cos^2\phi_1 +C_{k_2}^\dagger  D_{k_2}^\dagger A_{\tilde k_1}^\dagger  B_{\tilde k_1}^\dagger \sin^2\phi_1\cos^2\phi_1\\
+A_{k_1}^\dagger  D_{k_2}^\dagger C_{\tilde k_2}^\dagger  B_{\tilde k_1}^\dagger\sin^2\phi_1\cos^2\phi_1 -A_{k_1}^\dagger  D_{k_2}^\dagger A_{\tilde k_1}^\dagger  D_{\tilde k_2}^\dagger\cos^4\phi_1 \bigg]|0\rangle,
\label{part1}
\end{split}
\end{equation}

\begin{equation}\begin{split}
|\psi_{before BS}(4004)\rangle= \sum_{k \tilde k}\sqrt{\lambda_k\lambda_{\tilde k}}  
\sin^2\phi_1\cos^2\phi_1\bigg[A_{k_1}^\dagger  B_{k_1}^\dagger A_{\tilde k_1}^\dagger  B_{\tilde k_1}^\dagger+C_{k_2}^\dagger  D_{k_2}^\dagger C_{\tilde k_2}^\dagger  D_{\tilde k_2}^\dagger \bigg]|0\rangle,
\label{part2}
\end{split}
\end{equation}
\begin{equation}\begin{split}
|\psi_{before BS}(3113)\rangle= i\sum_{k \tilde k}\sqrt{\lambda_k\lambda_{\tilde k}}\\
\bigg[-C_{k_2}^\dagger  B_{k_1}^\dagger A_{\tilde k_1}^\dagger  B_{\tilde k_1}^\dagger\sin^3\phi_1\cos\phi_1 
+C_{k_1}^\dagger  B_{k_2}^\dagger C_{\tilde k_2}^\dagger  D_{\tilde k_2}^\dagger\sin\phi_1\cos^3\phi_1 \\
-A_{k_1}^\dagger  B_{k_1}^\dagger C_{\tilde k_2}^\dagger  B_{\tilde k_1}^\dagger\sin^3\phi_1\cos\phi_1
+C_{k_2}^\dagger  D_{k_2}^\dagger C_{\tilde k_1}^\dagger  B_{\tilde k_2}^\dagger\sin\phi_1\cos^3\phi_1 \\
+A_{k_1}^\dagger  D_{k_2}^\dagger A_{\tilde k_1}^\dagger  B_{\tilde k_1}^\dagger\sin\phi_1\cos^3\phi_1 
-A_{k_2}^\dagger  D_{k_1}^\dagger C_{\tilde k_2}^\dagger  D_{\tilde k_2}^\dagger\sin^3\phi_1\cos\phi_1 \\
+A_{k_1}^\dagger  B_{k_1}^\dagger A_{\tilde k_1}^\dagger  D_{\tilde k_2}^\dagger\sin\phi_1\cos^3\phi_1 
-C_{k_2}^\dagger  D_{k_2}^\dagger A_{\tilde k_2}^\dagger  D_{\tilde k_1}^\dagger\sin^3\phi_1\cos\phi_1\bigg]|0\rangle,
\label{part3}
\end{split}
\end{equation}

where the subscripts 1 and 2 indicate the first and the second spatial channels of the interferometer. The state in Eq.(\ref{total}) collects all scenarios that can be realized by the photons before the beam splitter. In particular, Eq.(\ref{part1}) describes the case with two photons in both channels, Eq.(\ref{part2}) corresponds to all photons in the same channel, whereas Eq.(\ref{part3}) indicates the situation with three photons in the first channel and one photon in the second channel (and vice versa). Contributions of these terms are basically related to the parameter $\phi_1$, if $\phi_1=0$ (or $\phi_1=\pi/2$), the most of the terms are vanished.

To analyze the degree of spatial entanglement between channels before the beam splitter, we calculate the reduced density matrix of the four-photon state before the beam splitter respect to the first channel and integrate this matrix over frequencies:

\begin{equation}\begin{split}
\rho_r=
\begin{bmatrix}
\frac{\cos^2 \phi_1}{4}(5-2\cos{2\phi_1}+\cos{4\phi_1}	)  &	A( \phi_1, \Delta l)\\
A( \phi_1, \Delta l)^*	&	\frac{\sin^2 \phi_1}{4}(5+2\cos{2\phi_1}+\cos{4\phi_1}	)
\end{bmatrix},
\end{split}\end{equation}

where the off-diagonal coefficients $A( \phi_1, \Delta l)$ have a tiny dependence on $\Delta l$, an explicit expression for $A( \phi_1, \Delta l)$ can be found in the Appendix A. 

\begin{figure}[H]
	\centering
	\includegraphics[width=0.8\linewidth]{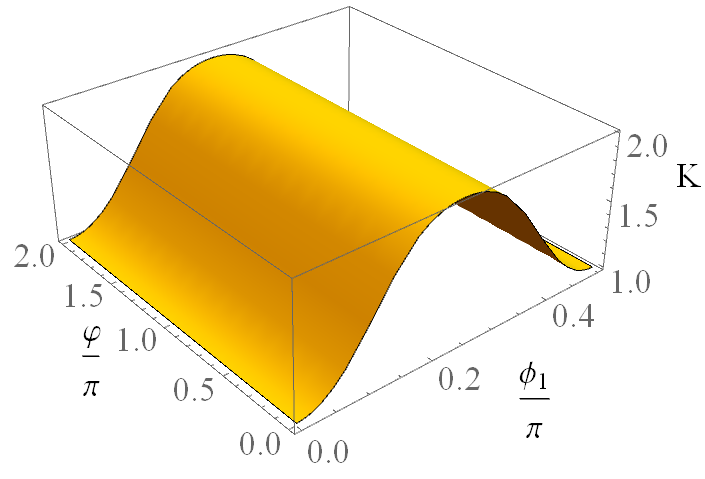}
	\caption{The Schmidt number K vs the conversion angle $\phi_1$ and the optical phase $ \varphi =\Delta L \ \omega_p/c$.}
	\label{Kfigure}
\end{figure}

The degree of spatial entanglement can be defined via introducing the Schmidt number $K=1/Tr(\rho^2_r)$.  For the zero path delay  $\Delta l=0$ the off-diagonal matrix elements $A( \phi_1, \Delta l)=0$ for any conversion parameter $\phi_1$, and the Schmidt number is reduced to:
\begin{equation}
K(\phi_1)=\frac{128}{94+33\cos{4 \phi_1} +2\cos{8\phi_1}-\cos{12 \phi_1}}.
\label{Schmidt}
\end{equation}
In the case $\phi_1=\pi/4$, the all wave functions, Eq.(\ref{part1}), Eq.(\ref{part2}) and Eq.(\ref{part3}), get the same weights which leads to maximizing the degree of entanglement for a bipartite system:  $K(\pi/4)=2$.
Meanwhile, for the fixed rotation angle $\phi_1=\pi/4$, the off-diagonal matrix elements are vanished for any path delay, and the Schmidt number achieves its maximum value $K=2$ for arbitary $\Delta l$. Fig.\ref{Kfigure} presents the Schmidt number in respect to the conversion angle $\phi_1$ and the optical phase  $ \varphi =\Delta l \ \omega_p/c$. It is clearly seen that the Schmidt number is slightly dependent on the path delay and achieves the maximum value with $\phi_1=\pi/4$. Together with spectral entaglement between the signal and idler photons, the generated spatial entanglement leads to a hyperentangled state.

\begin{figure}[H]
 	\centering
 	\includegraphics[width=0.8\linewidth]{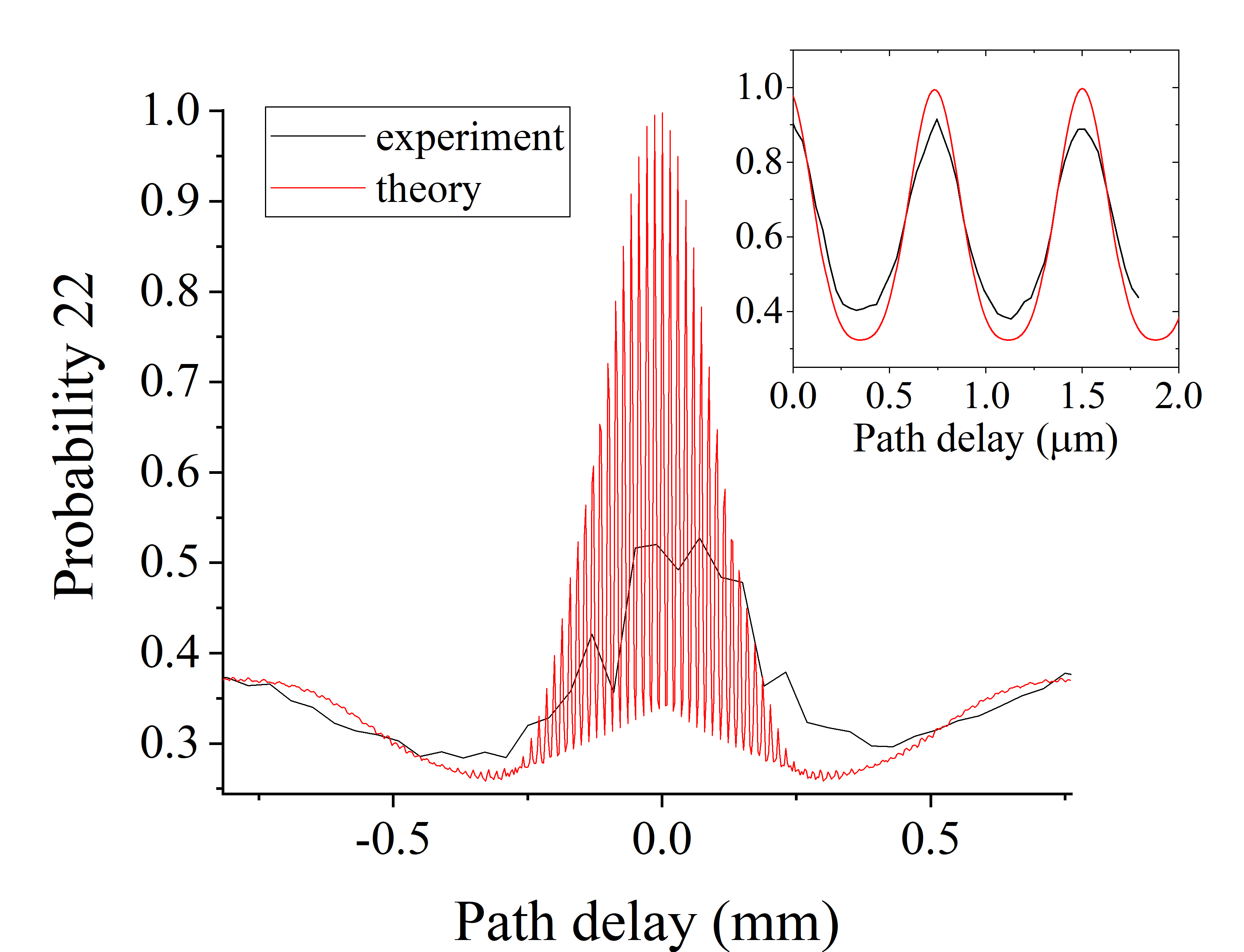}
 	\caption{The probability to detect two photons in each channel for the conversion angle $\phi_1=\pi/4$ in the single spectral mode regime. The red curve is theoretical calculations, while the black curve stands for the experimental data. The insert shows a zoom around zero time delay. The PDC section length is $L=8mm$, the pulse duration is $\tau=0.29 ps$, the pump wavelength is $\lambda_p=766nm$.}
 	\label{phi34}
 \end{figure}

With the use of the total unitary transformation of the state inside the interferometer, Eq.(\ref{utot}),  the coincidence probability to detect  photons in different channels can be calculated in terms of the positive-operator valued measures (POVM) \cite{PhysRevA.100.053829}. For example, the probability to detect two photons in each spatial channel, $P_{22}$, is shown in Fig.\ref{phi34}.

\begin{figure} [H]
	\centering
	\includegraphics[width=0.8\linewidth]{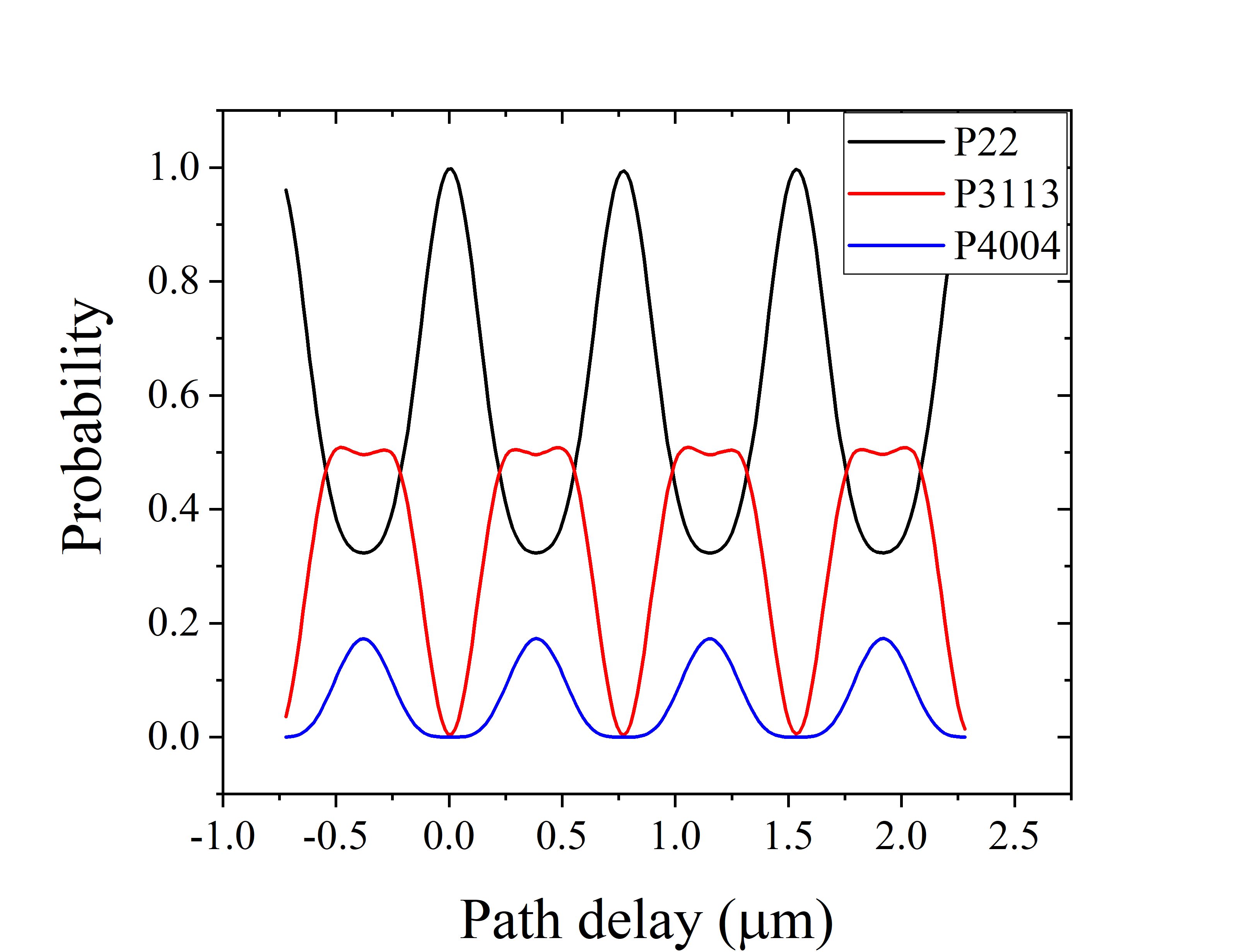}
	\caption{The zoom of the $P_{22}$, $P_{3113}$ and $P_{4004}$ probabilities around zero path delay for the parameters of Fig.\ref{phi34}. }
	\label{pnocomp}
\end{figure}

The $P_{22}$ probability has a complicated structure with two dips corresponding to the HOM interference at $\phi_1=0$ and $\phi_1=\pi/2$, the antibunching central peak arising due to the spatial entanglement, and very fast oscillations around zero path delay, similarly to the two-photon case \cite{PhysRevA.96.043857}.  The shape of the fast oscillations in the $P_{22}$ probability  is characterized by a more complicated structure respect to the two-photon case  \cite{PhysRevA.96.043857}, although the oscillation period still equals to the pump wavelength $\lambda_p$. In particular,  Eq.(\ref{part1}) provides the main trend of the curve and determines  the anti-bunching peak, two dips and a constant region, where the interference does not occur. Whereas Eq.(\ref{part2}) and Eq.(\ref{part3}) correspond to the oscillating terms in the $P_{22}$ probability with periods of oscillations $2\lambda_p$ and $\lambda_p$, respectively (for  more details, see Appendix A).  The zoom of the fast oscillations in cases $P_{3113}$ and $P_{4004}$ probabilities is shown in Fig. \ref{pnocomp}. It is clearly seen that the period of oscillations everywhere equals to the pump wavelength $\lambda_p$.



To prove our theoretical predictions we have conducted a series of experiments. As it can be seen from Fig.\ref{phi34}, experimental data give a good agreement with theoretical predictions for an envelope. The region in the center, around zero path delay, is characterized by very fast oscillations, see the zoom of Fig.\ref{phi34}, and needs more stable setup allowing to refine the detection precision and scan such interference fringes. With an improvement of the experimental setup, the agreement between theory and experiment becomes evident, see the zoom of Fig.\ref{phi34}.

\section{Experimental setup}

We use an engineered PDC source which can be configured to generate states with many speactral modes or nearly a single mode. The source is an 8 mm long periodically poled waveguide in potassium titanyl phosphate (PPKTP), with symmetric group velocity matching, which allows us to control the frequency correlation between signal and idler photons by shaping the pump pulses \cite{harder2013optimized, Ansari:18}. We pump the source with ultrashort pulses from a Ti:Sapph oscillator, followed by a pulse shaper. The pulse shaper is based on a spatial light modulator in a folded 4f setup and allows for shaping of both the spectral amplitude and phase of the pump pulses.

\begin{figure}[H]
	\centering
	\includegraphics[width=0.7\linewidth]{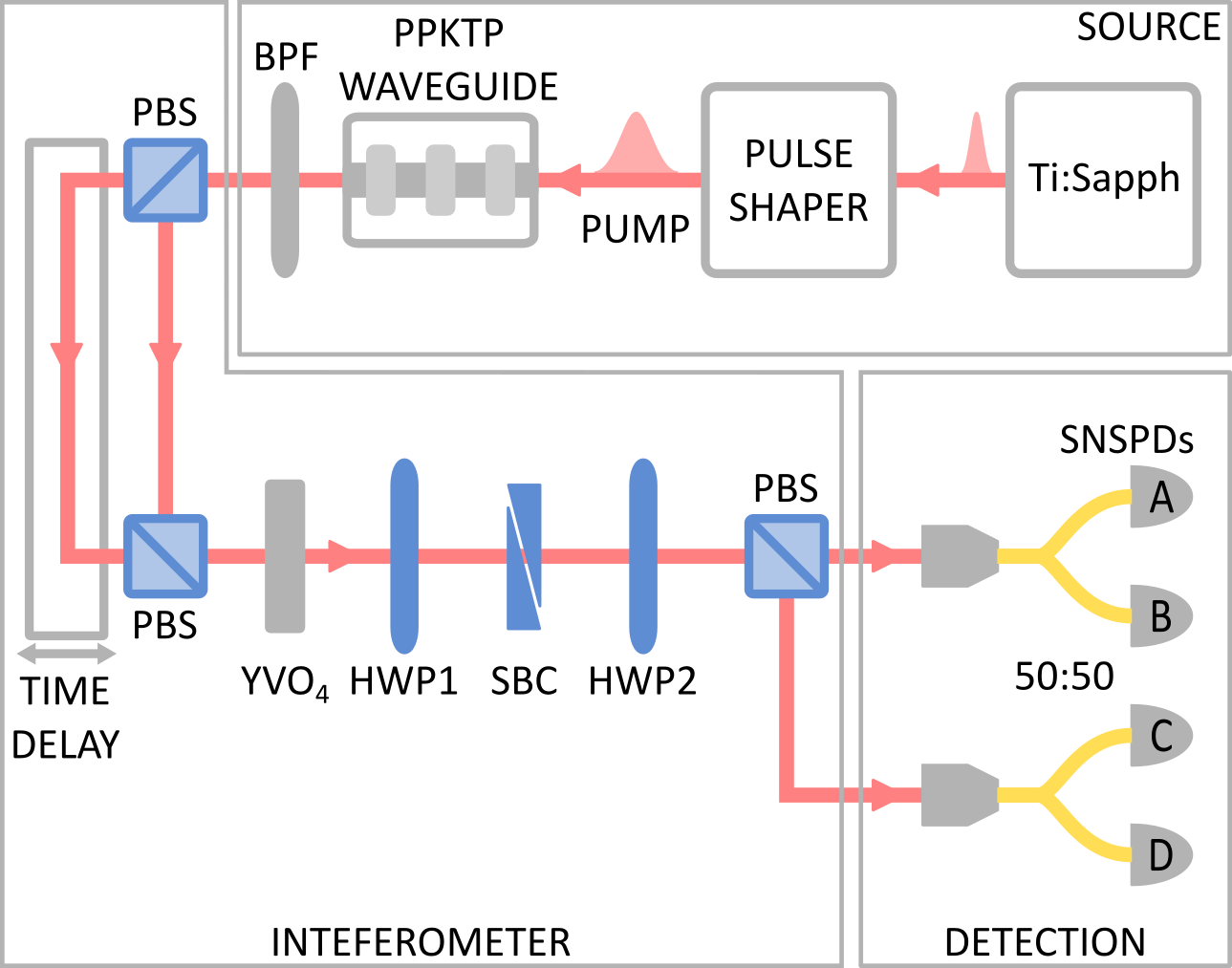}
	\caption{Experimental setup. A femtosecond titanium:sapphire (Ti:Sapph) oscillator with repetition rate of 80 MHz is used to pump a PPKTP waveguide designed for type-II PDC. We use a pulse shaper to set the desired spectral amplitude and phase of the pump pulses. An $8\,$nm wide bandpass filter (BPF) centered at $1532\,$nm was used to block the pump and phasematching side-lobes. The signal and idler photon coarse time delay is set with a linear translation stage. A $1.5\,$mm long YVO\textsubscript{4} crystal is used to compensate for the time delay between signal and idler photons, when the translation stage is set to zero delay. We use a common-path setup to interfere PDC photons where we scan the time delay using a soleil-babinet compensator (SBC). This is followed by half-wave plates $HWP_1$  and $HWP_2$, which are set to $22.5^\circ$, and a polarizing beamsplitter (PBS). This interferometer is equivalent to the setup shown in Fig. \ref{phim}. Photons are then coupled to balanced single-mode fiber splitters followed by four superconducting nanowire single photon detectors (SNSPD).}
	\label{exsetup}
\end{figure}

We use a phase-stable common-path setup to interfere the PDC photons, shown in Fig. \ref{exsetup}, which comprises half-waveplates, a soleil-babinet compensator, and a polarizing beam splitter. The interference pattern is scanned by adjusting the time delay on the soleil-babinet compensator. Note that the envelope in Fig. \ref{phi34} was measured by scaning the translation stage to adjust the coarse time delay between signal and idler photon. We used coherent light to test the interferometer.

\section{Bell states generation}
 Inside and outside the interferometer various combinations of Bell states can be generated. Since the size of the spectrum $\Delta \lambda$ and, therefore, the size of the Schmidt modes are much smaller than the wavelength of the signal/idler photon $\lambda_{s/i}$ (which corresponds to the period of oscillations of the function $ e^{i \omega_{s/i}\frac{\Delta l}{c}} $ regarding length),  i. e. $\Delta \lambda \ll \lambda_{s/i}$,  we can take the exponential function in Eq.(\ref{Schmidt_CD}) outside the integral:
 
\begin{equation}\begin{split}
C^\dagger_k \approx  e^{i \frac{\omega_p}{2} \frac{\Delta l}{ c}} \int d\omega_s u_k(\omega_s)  a^\dagger(\omega_s)= e^{i \frac{\omega_p}{2} \frac{\Delta l}{ c}} A^\dagger_k  ,\\
D^\dagger_k \approx  e^{i \frac{\omega_p}{2} \frac{\Delta l}{ c}} \int d\omega_i v_k(\omega_i)  a^\dagger(\omega_i)= e^{i \frac{\omega_p}{2} \frac{\Delta l}{ c}} B^\dagger_k .
\end{split}
\label{approx}
\end{equation} 
 

After interference of the state Eq. (\ref{total}) on the beam splitter with an angle $\theta$ under condition Eq. (\ref{approx}), the final output state can be written in the following form:
\begin{equation}\begin{split}
|\psi\rangle= \bigg[\sum_{k}\sqrt{\lambda_k}
\big(A_{k_2}^\dagger(-\cos\phi_1\sin\theta+e^{\pi i \Delta l/\lambda_p}\cos\theta\sin\phi_1)\\ +A_{k_1}^\dagger(-i\cos\theta\cos\phi_1-i e^{\pi i \Delta l/\lambda_p}\sin\theta\sin\phi_1)\big)\\
\big(B_{k_1}^\dagger(-i e^{\pi i \Delta l/\lambda_p}\cos\phi_1\sin\theta+i\cos\theta\sin\phi_1) \\
+B_{k_2}^\dagger(e^{\pi i \Delta l/\lambda_p}\cos\theta\cos\phi_1+\sin\theta\sin\phi_1)\big)\bigg]^2|0\rangle.
\label{free}
\end{split}\end{equation}
It is clearly seen that the output state strongly depends on the path delay. Modes $A$ and $B$ generated in the type-II PDC process have, in general, different spectral shapes and different group velocities because of different polarizations. Moreover, the number of generated modes can be different depending on the properties of the PDC source. 

For example, in the two-mode regime with $\lambda_1=1$, two sets of Schmidt modes exactly equals to each other $A=C$, $B=D$. Then for zero path delay $\Delta l=0$, Eq.(\ref{free}) gives :
\begin{equation}\begin{split}
|\psi_{\Delta l=0}\rangle= \bigg[ (A_{2}^\dagger\sin(\phi_1-\theta)-i A_{1}^\dagger\cos(\phi_1-\theta)) (i B_{1}^\dagger\sin(\phi_1-\theta)+ B_{2}^\dagger\cos(\phi_1-\theta))\bigg]^2|0\rangle.
\end{split}\end{equation}
The output state in such case always depends on the difference $(\phi_1-\theta)$ between the polarization rotation $\phi_1$ and beam splitter  $\theta$ angles. When $\phi_1=\theta$, the output state is $\big(A_1^\dagger B_2^\dagger\big)^2|0\rangle$, which corresponds to the state having two spectrally distinguishable photon pairs in two different channels. 
In the case with fixing $\phi_1-\theta=\pi/4$ we can write the output state in terms of the spectral bidimensional basis A and B:
\begin{equation}\begin{split}
|\psi_{\Delta l=0}\rangle=\bigg[\frac{i}{2}(|g\rangle_A|r\rangle_B-|r\rangle_A|g\rangle_B)+\frac{1}{2}(|r\rangle_A|r\rangle_B+|g\rangle_A|g\rangle_B)\bigg]^2 \\
=\frac{1}{2}(-i|\Psi^-\rangle +|\Phi^+\rangle)^2,
\label{bell1}
\end{split}\end{equation}
where $r$ and $g$ are the red (first) and the green (second) spatial channels in Fig.(\ref{phim}) respectively. This state can be rewritten in terms of the Bell states $\Psi^-$ and $\Phi^+$ from the Bell state basis defined in our notation as following:
\begin{equation}\begin{split}
|\Psi^+\rangle=\frac{1}{\sqrt 2}(|r\rangle_A|g\rangle_B+|g\rangle_A|r\rangle_B)\\
|\Psi^-\rangle=\frac{1}{\sqrt 2}(|r\rangle_A|g\rangle_B-|g\rangle_A|r\rangle_B)\\
|\Phi^+\rangle=\frac{1}{\sqrt 2}(|r\rangle_A|r\rangle_B+|g\rangle_A|g\rangle_B)\\
|\Phi^-\rangle=\frac{1}{\sqrt 2}(|r\rangle_A|r\rangle_B-|g\rangle_A|g\rangle_B).
\end{split}\end{equation}
By fixing $\Delta l=\lambda_p$ in Eq.(\ref{free}) in the two-mode regime we have the following state:
\begin{equation}\begin{split}
|\psi_{\Delta l=\lambda_p}\rangle= \bigg[\big(-A_{2}^\dagger\sin(\phi_1+\theta)-i A_{1}^\dagger\cos(\phi_1+\theta)\big) \\
\big(i B_{1}^\dagger\sin(\phi_1+\theta)- B_{2}^\dagger\cos(\phi_1+\theta)\big)\bigg]^2|0\rangle.
\end{split}\end{equation}
When $\phi_1+\theta=\pi/4$, this state gives the following Bell states combination:  
\begin{equation}\begin{split}
|\psi_{\Delta l=\lambda_p}\rangle=\frac{1}{2}(|\Phi^+\rangle + i|\Psi^-\rangle)^2
\label{bell2}
\end{split}\end{equation}
It is relevant to observe that Eq.(\ref{bell1}) and Eq.(\ref{bell2}) can be both achieved in the same interferometer by fixing $\phi_1=\pi/4$ and $\theta=0$ (without the beam splitter) and varying the path delay. 

Keeping parameters $\phi_1=\pi/4$ and $\theta=0$ and making the same procedure as before, we can consider two further cases, $\Delta l=\lambda_p/2$ and $\Delta l=3\lambda_p/2$, which provide the following states respectively:
\begin{equation}\begin{split}
|\psi_{\Delta l=\frac{\lambda_p}{2}}\rangle=\frac{1}{2}(|\Phi^-\rangle + |\Psi^-\rangle)^2,
\end{split}\end{equation}
\begin{equation}\begin{split}
|\psi_{\Delta l=\frac{3\lambda_p}{2}}\rangle=\frac{1}{2}(|\Phi^-\rangle - |\Psi^-\rangle)^2.
\end{split}\end{equation}

All the states which were described above can be represented through the four-qubit Bell gem $ \left\lbrace G_i \right\rbrace  $, see Appendix, introduced in \cite{JAEGER2004425, Jaeger2008}. For example, the state in Eq.(\ref{bell1}) can be written in the following form:
\begin{equation}\begin{split}
|\psi\rangle=\frac{1}{2}\big( (|\Phi^+\rangle |\Phi^+\rangle -|\Psi^-\rangle |\Psi^-\rangle) - i(|\Phi^+\rangle |\Psi^-\rangle +|\Psi^-\rangle |\Phi^+\rangle)\big) \\ 
=  \frac{1}{\sqrt 2}(G_1^+ +G_1^- +G_2^- - G_2^+) -  \frac{i}{\sqrt 2}\,G_5^+,
\label{ent}
\end{split}\end{equation}
The  combination of the  gem states in the bracket corresponds to the even number of photons in the channels, while the last term $G_5$, nothing more then $|\psi (3113)\rangle$, namely, Eq.(7). It means that with a post-selection, the  high-entangled four-dimensional Bell state with the odd nimber of photons in the channel can be generated in the presented setup.

\section{Manipulation of fast oscillations}
In this section we consider identical Schmidt mode profiles of the signal and idler photons, which means the same spectral shapes and the same phases of both photons.

To realize this situation, firstly, we increase the number of spectral modes via enhancing the pulse duration of the pump laser. Indeed, in such case we ensure the spectral indistinguishability of the signal-idler photons, namely, $A_n $ and $B_n$ become identical amplitudes. Secondly, we have to compensate the phase between the signal and idler photons. In the first order of decomposition, such phase corresponds to the time delay between photons. This time delay can be compensated by adding the polarization converter after the PDC source which perfectly converts the polarization of both signal and idler photons and a further non-poled KTP crystal having length $L_{PDC}/2$. Since the signal and idler photon pairs are generated effectively in the middle of the PDC section, such new configuration allows to compensate perfectly the arising time delay.

\begin{figure}[H]
	\centering
	\includegraphics[width=0.8\linewidth]{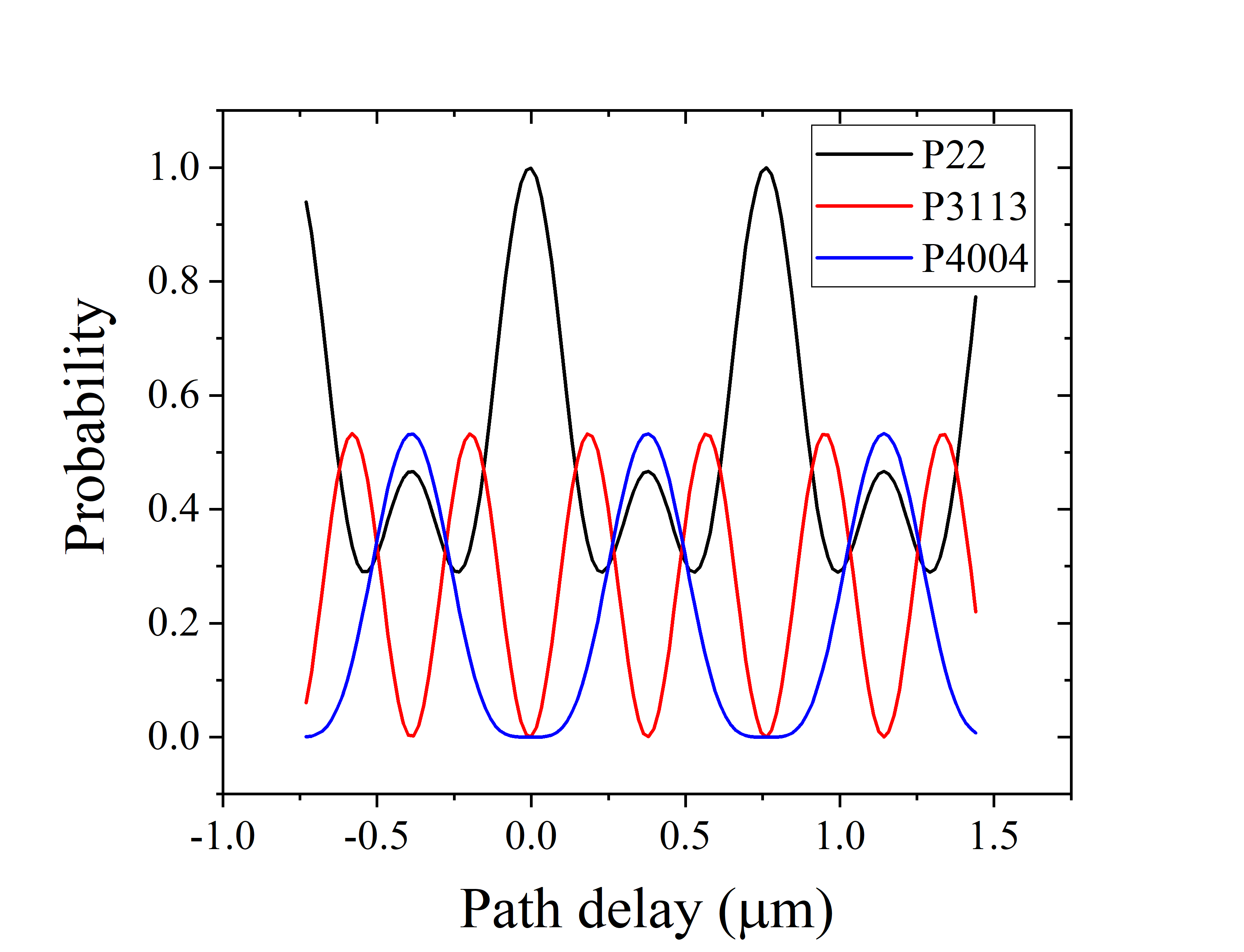}
	\caption{The zoom of the $P_{22}$, $P_{3113}$ and $P_{4004}$  probabilities in the configuration described by the identical signal-idler Schmidt operators. Pulse duration $\tau=10 ps$.}
	\label{pcomp}
\end{figure}

Mathematically this modification can be done 
by adding a proper phase in the JSA in Eq.(\ref{jsa}). The new JSA gives
\begin{equation}
F(\omega_s,\omega_i)=e^{-(\omega_s+\omega_i-\omega_p)\tau^2/2}\sinc\bigg(\frac{L}{2}\Delta k\bigg)e^{i \frac{L}{2}(\Delta k+\bar{k})},
\end{equation}
where $\bar{ k}=-k_e(\omega_s)-k_o(\omega_i)$ indicates signal and idler photons having opposite polarizations respect to $\Delta k$. In this case the Schmidt modes of the signal and idler photons equals to each other $A_k=B_k$. From the general output state Eq.($\ref{free}$) under the most interesting condition $\phi_1=\theta=\pi/4$ one can obtain the wave function in the end of the interferometer: 
\begin{equation}
|\psi\rangle=|\psi(22)\rangle+|\psi(4004)\rangle+|\psi(3113)\rangle,
\end{equation}
where
\begin{equation}\begin{split}
|\psi(22)\rangle=-\frac{e^{2\pi i \Delta l/\lambda_p}}{4}\sum_{k \tilde k}\sqrt{\lambda_k\lambda_{\tilde k}}\\
\bigg[\sin^2\bigg(\frac{\pi \Delta l}{\lambda_p}\bigg)\big({A_{k_1}^\dagger}^2 {A_{\tilde k_2}^\dagger}^2+{A_{\tilde k_1}^\dagger}^2{A_{k_2}^\dagger}^2 \big)+4\cos^2\bigg(\frac{\pi \Delta l}{\lambda_p}\bigg) A_{k_1}^\dagger A_{\tilde k_1}^\dagger A_{k_2}^\dagger A_{\tilde k_2}^\dagger\bigg]|0\rangle,
\end{split}\end{equation}
\begin{equation}\begin{split}
|\psi(4004)\rangle=- \frac{e^{2\pi i \Delta l/\lambda_p}}{4}\sum_{k \tilde k}\sqrt{\lambda_k\lambda_{\tilde k}}\sin^2\bigg(\frac{\pi \Delta l}{\lambda_p}\bigg)\bigg({A_{k_2}^\dagger}^2 {A_{\tilde k_2}^\dagger}^2+ {A_{k_1}^\dagger}^2  {A_{\tilde k_1}^\dagger}^2\bigg)|0\rangle,
\end{split}\end{equation}
\begin{equation}\begin{split}
|\psi(3113)\rangle= -\frac{e^{2\pi i \Delta l/\lambda_p}}{4}\bigg[\sum_{k \tilde k}\sqrt{\lambda_k\lambda_{\tilde k}}\sin\bigg(\frac{2\pi \Delta l}{\lambda_p}\bigg)\\
\big({A_{k_1}^\dagger}^2  A_{\tilde k_1}^\dagger A_{\tilde k_2}^\dagger+A_{k_1}^\dagger  A_{k_2}^\dagger{A_{\tilde k_1}^\dagger}^2  +{A_{k_2}^\dagger}^2 A_{\tilde k_2}^\dagger A_{\tilde k_1}^\dagger+ {A_{\tilde k_2}^\dagger}^2 A_{ k_2}^\dagger A_{k_1}^\dagger 
\big)|0\rangle.
\end{split}\end{equation}
The coincidence probability around zero path delay for the state written above is presented in Fig.\ref{pcomp}. It is clearly seen that the period of $P_{22}$ and $P_{4004}$ oscillations equals to $\lambda_p$. Of course there are small peaks in $P_{22}$ probability that can give the period of oscillations $\lambda_p/2$ together with high peaks but the visibility is quite low in this case.  Simultaneously, the period of oscillations in $P_{3113}$ probability is twice smaller: $\lambda_p/2$ with a high visibility. This fact can be used to increase the sensitivity of measurements by changing only the coincidence scheme.

Moreover, the peaks in $P_{22}$ close to the unity correspond to the  path delay  $\Delta l= m\lambda_p$, where $m$ are integer numbers. The maximum of $P_{4004}$ is given by the path delay $\Delta l=(2m+1)\lambda_p/2$. Lastly, the peaks in $P_{3113}$ are given by the path delay $\Delta l=(m+1) \lambda_p/4$. 
As expected, more symmetrical JSA along the diagonal frequency axis, namely more similar the signal and idler photon spectra, the closer the peaks of $P_{22}$ and $P_{4004}$ when $\Delta l=\lambda_p/2$, which is drastically different from  Fig.\ref{pnocomp}. Therefore, the output state in the case of identical Schmidt modes of the signal and idler photons with a fixed path delay $\Delta l=\lambda_p/2$ in the interferometer  can be  described by $|\psi\rangle=(|40\rangle+|04\rangle)/4+|22\rangle/4$, which is a balanced superposition between a NOON state and the input state of the interferometer.
\section{Conclusion}
We have presented a modified scheme of the four-photon Hong-Ou-Mandel interferometer based on a type-II PDC process with an additional polarization converter which creates a polarization superposition for both signal and idler photons. By fixing properly the conversion parameter, we maximized the degree of entanglement between two spatial channels of the interferometer. Moreover, the presented device allows to generate fast oscillations in the interference pattern. In contrast to the already known two photon case, this fast oscillations present a more complex structure due to an additional interference processes, which lead to the period of fast oscillations twice smaller than the pump wavelength in the  regime of indistinguishable Schmidt modes. In addition, a balanced superposition between the initial state and the NOON state can be generated. Moreover, the output radiation was studied in terms of Bell-states in the Schmidt modes basis. A generation of various combinations of the four-qubit Bell states by properly manipulating both the conversion and the beam splitter transmission/reflection parameters was shown. Several of such states can be conveniently isolated via post-selecting strategies and used in following operations.  Our work can be used in future studies as a starting point for generating further high-entangled Bell states and for improving the measurement precision strategy.
\section{ACKNOWLEDGMENTS}
We acknowledge the financial support of the Deutsche Forschungsgemeinschaft (DFG)
via TRR~142/2, project C02 and the project SH 1228/3-1.  P.~R.~Sh. thanks the state of North Rhine-Westphalia for support by the
{\it Landesprogramm f{\"u}r geschlechtergerechte Hochschulen}.
\section*{References}
\bibliographystyle{unsrt}
\bibliography{database}

\appendix

\section{Four-photon state generated in the interferometer}

The input photons of the interferometer are generated by a type-II PDC process. Mathematically, the Hamiltonian which describes this process is \cite{quesada2014effects}:
\begin{equation}
H=i \hbar \Gamma \int \mathsf{d}\omega_s \mathsf{d}\omega_i  F(\omega_s,\omega_i) a_H^\dagger(\omega_s) a_V^\dagger(\omega_i) + h.c.
\end{equation}
with the joint spectral amplitude (JSA)
\begin{equation}
F(\omega_s,\omega_i)=e^{-(\omega_s+\omega_i-\omega_p)\tau^2/2}\sinc\bigg(\frac{L}{2}\Delta k\bigg)e^{i \frac{L}{2}\Delta k}
\label{jsa}
\end{equation}
 where $p$, $s$ and $i$ indicate respectively the pump, signal and idler photons, $\Gamma$ is the coupling constant, which contains information about the PDC length, the pulse duration and pump intensity. The function $\Delta k=k_o(\omega_p)-k_o(\omega_s)-k_e(\omega_i)+2\pi/\Lambda$ is the phase mismatch which represents the momentum conservation of the process, $\Lambda=126\mu m$ is the poling period.

In a low gain regime, which means that $\langle N \rangle \ll 1$ where N is the number of generated photons,  the four-photon state can be calculated by using the second-order perturbation theory without taking into account the time ordering effect as follows:
\begin{equation}\begin{split}
|\psi_{4 ph}\rangle = 
\frac{1}{2}\bigg( \int_0^t H\mathsf{d}t'\bigg)^2 \mid 0 \rangle = 
\frac{\xi^2 }{2}\int \mathsf{d}\omega_s\mathsf{d}\omega_i  F(\omega_s,\omega_i)
a_H^\dagger(\omega_s)a_V^\dagger(\omega_i)\\ 
\int \mathsf{d}\tilde\omega_s \mathsf{d}\tilde\omega_i F(\tilde\omega_s,\tilde\omega_i) a_H^\dagger(\tilde\omega_s)a_V^\dagger(\tilde\omega_i)|0\rangle, \ \
\label{pdc}
\end{split}\end{equation} 
where $\xi = \Gamma t$, $H$, $V$ label the horizontal and vertical polarization respectievely. The wavefunction before the beam splitter can be written with acting of the unitary transformation 
\begin{equation}
U_{before BS} = FP_3\cdot HWP_2\cdot FP_2\cdot PBS\cdot FP_1\cdot HWP_1\cdot FP_0
\label{ubef}
\end{equation}
on the input wave function Eq.(\ref{pdc}). The wave function before the beam splitter depends on the distances between the optical elements ($x$, $y$, $z$), the path delay $\Delta l$ and the conversion angle $\phi_1$ of the first polarization converter:

\begin{equation}\begin{split}
|\psi_{before BS}(22)\rangle= \int \mathsf{d}\omega_s \mathsf{d}\omega_i  \mathsf{d}\tilde\omega_s \mathsf{d}\tilde\omega_i F(\omega_s,\omega_i)F(\tilde\omega_s,\tilde\omega_i) e^{i(\omega_s+\omega_i+\tilde\omega_s+\tilde\omega_i)(x+y+l)/c} \\  
\bigg[-a^\dagger_2(\omega_s)a^\dagger_1(\omega_i)a^\dagger_2(\tilde\omega_s)a^\dagger_1(\tilde\omega_i)\sin^4\phi_1 e^{i(\omega_s+\tilde\omega_s)\frac{\Delta l}{c}}\\
+a^\dagger_2(\omega_s)a^\dagger_1(\omega_i)a^\dagger_1(\tilde\omega_s)a^\dagger_2(\tilde\omega_i) \sin^2\phi_1\cos^2\phi_1 e^{i(\omega_s+\tilde\omega_i)\frac{\Delta l}{c}}\\
+a^\dagger_1(\omega_s)a^\dagger_1(\omega_i)a^\dagger_2(\tilde\omega_s)a^\dagger_2(\tilde\omega_i)\sin^2\phi_1\cos^2\phi_1 e^{i(\tilde\omega_s+\tilde\omega_i)\frac{\Delta l}{c}}\\
+a^\dagger_2(\omega_s)a^\dagger_2(\omega_i)a^\dagger_1(\tilde\omega_s)a^\dagger_1(\tilde\omega_i)\sin^2\phi_1\cos^2\phi_1 e^{i(\omega_s+\omega_i)\frac{\Delta l}{c}}\\
+a^\dagger_1(\omega_s)a^\dagger_2(\omega_i)a^\dagger_2(\tilde\omega_s)a^\dagger_1(\tilde\omega_i)\sin^2\phi_1\cos^2\phi_1 e^{i(\omega_i+\tilde\omega_s)\frac{\Delta l}{c}}\\
-a^\dagger_1(\omega_s)a^\dagger_2(\omega_i)a^\dagger_1(\tilde\omega_s)a^\dagger_2(\tilde\omega_i)\cos^4\phi_1 e^{i(\omega_i+\tilde\omega_i)\frac{\Delta l}{c}}\bigg]|0\rangle,
\label{part11}
\end{split}\end{equation}
\begin{equation}\begin{split}
|\psi_{before BS}(4004)\rangle= \int \mathsf{d}\omega_s \mathsf{d}\omega_i  \mathsf{d}\tilde\omega_s \mathsf{d}\tilde\omega_i F(\omega_s,\omega_i)F(\tilde\omega_s,\tilde\omega_i) e^{i(\omega_s+\omega_i+\tilde\omega_s+\tilde\omega_i)(x+y+l)/c} \\  
\frac{\sin^22\phi_1}{4}\bigg[a^\dagger_1(\omega_s)a^\dagger_1(\omega_i)a^\dagger_1(\tilde\omega_s)a^\dagger_1(\tilde\omega_i)+a^\dagger_2(\omega_s)a^\dagger_2(\omega_i)a^\dagger_2(\tilde\omega_s)a^\dagger_2(\tilde\omega_i)e^{i(\omega_s+\omega_i+\tilde\omega_s+\tilde\omega_i)\frac{\Delta l}{c}}\bigg]|0\rangle,
\label{part22}
\end{split}\end{equation}
\begin{align}\begin{split}
|\psi_{before BS}(3113)\rangle= i\int \mathsf{d}\omega_s \mathsf{d}\omega_i  \mathsf{d}\tilde\omega_s \mathsf{d}\tilde\omega_i F(\omega_s,\omega_i)F(\tilde\omega_s,\tilde\omega_i) e^{i(\omega_s+\omega_i+\tilde\omega_s+\tilde\omega_i)(x+y+l)/c} \\
\bigg[-a^\dagger_2(\omega_s)a^\dagger_1(\omega_i)a^\dagger_1(\tilde\omega_s)a^\dagger_1(\tilde\omega_i)\sin^3\phi_1\cos\phi_1 e^{i \omega_s\frac{\Delta l}{c}}\\
+a^\dagger_1(\omega_s)a^\dagger_2(\omega_i)a^\dagger_2(\tilde\omega_s)a^\dagger_2(\tilde\omega_i)\sin\phi_1\cos^3\phi_1 e^{i (\tilde\omega_s+\tilde\omega_i+\omega_s)\frac{\Delta l}{c}} \\
-a^\dagger_1(\omega_s)a^\dagger_1(\omega_i)a^\dagger_2(\tilde\omega_s)a^\dagger_1(\tilde\omega_i)\sin^3\phi_1\cos\phi_1 e^{i \tilde\omega_s\frac{\Delta l}{c}}\\
+a^\dagger_2(\omega_s)a^\dagger_2(\omega_i)a^\dagger_1(\tilde\omega_s)a^\dagger_2(\tilde\omega_i)\sin\phi_1\cos^3\phi_1 e^{i (\omega_s+\omega_i+\tilde\omega_s)\frac{\Delta l}{c}}\\
+a^\dagger_1(\omega_s)a^\dagger_2(\omega_i)a^\dagger_1(\tilde\omega_s)a^\dagger_1(\tilde\omega_i)\sin\phi_1\cos^3\phi_1 e^{i \omega_i\frac{\Delta l}{c}}\\
-a^\dagger_2(\omega_s)a^\dagger_1(\omega_i)a^\dagger_2(\tilde\omega_s)a^\dagger_2(\tilde\omega_i)\sin^3\phi_1\cos\phi_1 e^{i (\tilde\omega_s+\tilde\omega_i+\omega_i)\frac{\Delta l}{c}}\\
+a^\dagger_1(\omega_s)a^\dagger_1(\omega_i)a^\dagger_1(\tilde\omega_s)a^\dagger_2(\tilde\omega_i)\sin\phi_1\cos^3\phi_1 e^{i \tilde\omega_i\frac{\Delta l}{c}}\\
-a^\dagger_2(\omega_s)a^\dagger_2(\omega_i)a^\dagger_2(\tilde\omega_s)a^\dagger_1(\tilde\omega_i)\sin^3\phi_1\cos\phi_1 e^{i (\omega_s+\omega_i+\tilde\omega_i)\frac{\Delta l}{c}}\bigg]|0\rangle,
\label{part33}
\end{split}\end{align}
where $c$ is the speed of light.
From this expression it is possible to calculate the reduced density matrix, Eq.(\ref{total}), where
\begin{equation}\begin{split}
A( \phi_1, \Delta l)=\int \mathsf{d}\omega_s \mathsf{d}\omega_i  \mathsf{d}\tilde\omega_s \mathsf{d}\tilde\omega_i \mid F(\omega_s,\omega_i)\mid^2\mid F(\tilde\omega_s,\tilde\omega_i)\mid^2 \times \\
\bigg[\sin^4{\phi_1}\cos^4\phi_1 (e^{-i(\tilde\omega_s+\tilde\omega_i)\frac{\Delta l}{c}}+e^{-i(\omega_s+\omega_i)\frac{\Delta l}{c}})\\
-\sin^2\phi_1\cos^6\phi_1 e^{-i(\tilde\omega_s+\omega_i)\frac{\Delta l}{c}}-\sin^6\phi_1\cos^2\phi_1 e^{-i(\omega_s+\tilde\omega_i)\frac{\Delta l}{c}} \\
-i\sin^5\phi_1 \cos^3\phi_1 e^{-i(\omega_s+\omega_i-\tilde\omega_i)\frac{\Delta l}{c}}+i \sin^7\phi_1\cos\phi_1 e^{-i\tilde\omega_s\frac{\Delta l}{c}}
\\
+i\sin^3\phi_1\cos^5\phi_1 e^{-i(\tilde\omega_s+\tilde\omega_i-\omega_i)\frac{\Delta l}{c}}-i \sin\phi_1\cos^7\phi_1 e^{-i\omega_s\frac{\Delta l}{c}}\bigg],
\end{split}\end{equation}
and to get the Schmidt number given in Eq.(\ref{Schmidt}) and plotted in Fig.\ref{Kfigure}.
Moreover, Eqs. (\ref{part11},\ref{part22},\ref{part33}) can be simplified via introducing the Schmidt decomposition of the JSA $F(\omega_s, \omega_i)=\sum_{k}\sqrt{\lambda_k} \tilde{u}_k(\omega_s)\tilde{v}_k(\omega_i)$ and two sets of the broadband operators: 

path-independent
\begin{equation}\begin{split}
A^\dagger_k=\int d\omega_s u_k(\omega_s)a^\dagger(\omega_s), \\
B^\dagger_k=\int d\omega_i v_k(\omega_i)a^\dagger(\omega_i),
\label{AB}
\end{split}
\end{equation} 
and path-dependent
\begin{equation}\begin{split}
C^\dagger_k=\int d\omega_s u_k(\omega_s)e^{i \omega_s\frac{\Delta l}{c}}   a^\dagger(\omega_s) ,\\
D^\dagger_k=\int d\omega_i v_k(\omega_i) e^{i \omega_i\frac{\Delta l}{c}}  a^\dagger(\omega_i),
\label{CD}
\end{split}
\end{equation} 
where $u_k(\omega_s)= \tilde{u}_k(\omega_s)e^{i \omega_s (x+y+l)/c}$ and $v_k(\omega_i)= \tilde{v}_k(\omega_i)e^{i \omega_i (x+y+l)/c}$.  Eqs. (\ref{part11},\ref{part22},\ref{part33}) can be rewritten in terms of broadband operators Eq.(\ref{AB}, \ref{CD}) in the form of Eq.(\ref{part1}, \ref{part2}, \ref{part3}) in the main text.

The output state in the  end of the interferometer with rotation angles of $HWP_1$ and $BS$ equal to $\phi_1$ and $\theta$, respectively, is given by

\begin{equation}\begin{split}
	|\psi_{out}\rangle= \int \mathsf{d}\omega_s \mathsf{d}\omega_i  \mathsf{d}\tilde\omega_s \mathsf{d}\tilde\omega_i F(\omega_s,\omega_i)F(\tilde\omega_s,\tilde\omega_i) e^{i(\omega_s+\omega_i+\tilde\omega_s+\tilde\omega_i)(x+y+l)/c} \\  
	[a_2^\dagger(\omega_s)(-\cos\phi_1\sin\theta+e^{i\omega_s \Delta l/c}\cos\theta\sin\phi_1)+ a_1^\dagger(\omega_s)(-i\cos\theta\cos\phi_1-i e^{i\omega_s \Delta l/c}\sin\theta\sin\phi_1)]\\
	[ a_1^\dagger(\omega_i)(-i e^{i\omega_i \Delta l/c}\cos\phi_1\sin\theta+i\cos\theta\sin\phi_1)+a_2^\dagger(\omega_i)(e^{i\omega_i \Delta l/c}\cos\theta\cos\phi_1+\sin\theta\sin\phi_1)]\\
	[a_2^\dagger(\tilde\omega_s)(-\cos\phi_1\sin\theta+e^{i\tilde\omega_s \Delta l/c}\cos\theta\sin\phi_1)+ a_1^\dagger(\tilde\omega_s)(-i\cos\theta\cos\phi_1-i e^{i\tilde\omega_s \Delta l/c}\sin\theta\sin\phi_1)]\\
[ a_1^\dagger(\tilde\omega_i)(-i e^{i\tilde\omega_i \Delta l/c}\cos\phi_1\sin\theta+i\cos\theta\sin\phi_1)+a_2^\dagger(\tilde\omega_i)(e^{i\tilde\omega_i \Delta l/c}\cos\theta\cos\phi_1+\sin\theta\sin\phi_1)].
	\label{free1}
	\end{split}\end{equation}
	

 By varying opportunely all $\theta$,$\phi$ and $\Delta l$, it is possible to generate different outputs of the four-qubit Bell gem $\{G_i\}$ introduced in \cite{JAEGER2004425}:

\begin{equation}\begin{split}
G_1^\pm=\frac{1}{\sqrt 2}(|\Phi^+\rangle|\Phi^+\rangle\pm |\Phi^-\rangle|\Phi^-\rangle),\\
G_2^\pm=\frac{1}{\sqrt 2}(|\Psi^+\rangle|\Psi^+\rangle\pm |\Psi^-\rangle|\Psi^-\rangle) \\
G_3^\pm=\frac{1}{\sqrt 2}(|\Phi^+\rangle|\Phi^-\rangle\pm |\Phi^-\rangle|\Phi^+\rangle),\\
G_4^\pm=\frac{1}{\sqrt 2}(|\Phi^+\rangle|\Psi^+\rangle\pm |\Psi^+\rangle|\Phi^+\rangle),\\
G_5^\pm=\frac{1}{\sqrt 2}(|\Phi^+\rangle|\Psi^-\rangle\pm |\Psi^-\rangle|\Phi^+\rangle),\\
G_6^\pm=\frac{1}{\sqrt 2}(|\Phi^-\rangle|\Psi^+\rangle\pm |\Psi^+\rangle|\Phi^-\rangle),\\
G_7^\pm=\frac{1}{\sqrt 2}(|\Phi^-\rangle|\Psi^-\rangle\pm |\Psi^-\rangle|\Phi^-\rangle),\\
G_8^\pm=\frac{1}{\sqrt 2}(|\Psi^+\rangle|\Psi^-\rangle\pm |\Psi^-\rangle|\Psi^+\rangle),
\end{split}\end{equation}
which is discussed in Section IV.

The expression which represents the curve in Fig.\ref{phi34} is therefore given by Eq.(\ref{free1}) fixing $\theta=\phi_1=\pi/4$ and taking the only relative terms describing two photons in two different channels. As was discussed early, the main trend of the curve in Fig.\ref{phi34} is due to the term in Eq.(\ref{part11}), namely it determines the central peak and the sidedips as well as the constant region, where the interference does not occur. The fast oscillations happen due to the terms in Eq.(\ref{part22}) and Eq.(\ref{part33}). 

Of course, all the terms Eq.(\ref{part11}), Eq.(\ref{part22})  and Eq.(\ref{part33}) interfere at the beam splitter before measuring of the coincidence probability. But it is interesting to note that terms with the even and the odd number of photons in the channels, for example, Eq.(\ref{part11}) and Eq.(\ref{part33}) or  Eq.(\ref{part22}) and Eq.(\ref{part33}) do not interfere at the beam splitter. It means that there is no contribution to the coincidence probability from the interference at the beam splitter between terms with the even and the odd number of photons.  Only terms with the even number of photons, Eq.(\ref{part11}) and Eq.(\ref{part22}), interfere at the beam splitter  due to the bosonic nature of light.

\begin{figure*}[h]
	\includegraphics[width=1\linewidth]{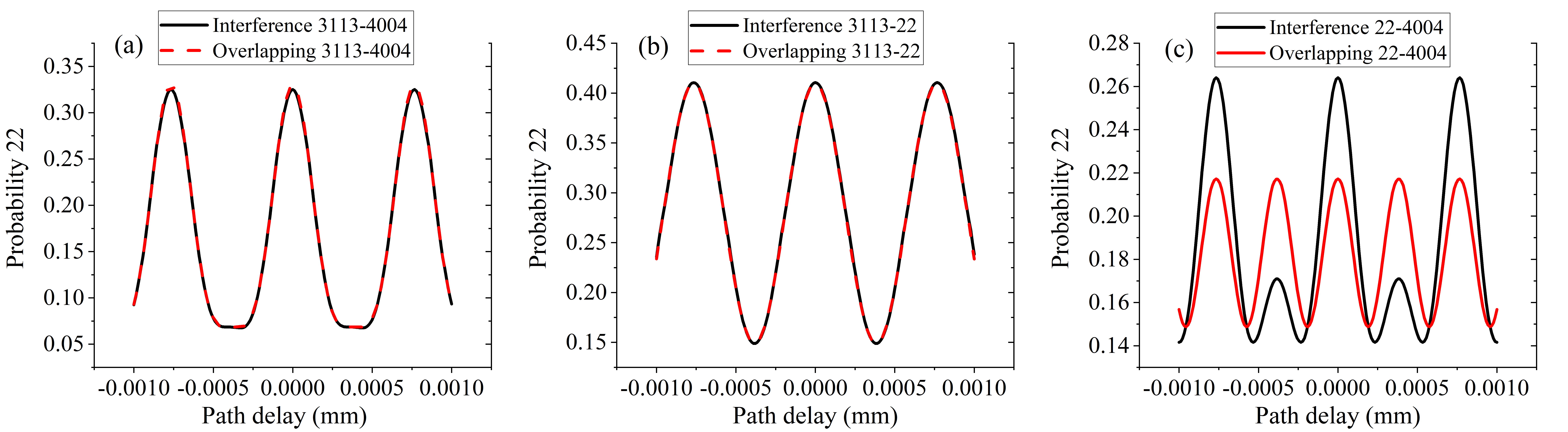}
	\caption{A comparison between the "overlapping terms" and the "interference terms" from Eq(\ref{part1}), Eq(\ref{part2}) and Eq(\ref{part3}) calculated around the zero-path delay.}
	\label{oscill}
\end{figure*}

In order to make this point explicit, in Fig.\ref{oscill} we distinguish the "interference terms", namely the terms obtained by letting Eq.(\ref{part11}), Eq.(\ref{part22}) and Eq.(\ref{part33}) interfere in pairs at the beam splitter, from the "overlapping terms", namely the terms we got if we would calculate the coincidence probabilities from Eq.(\ref{part11}), Eq.(\ref{part22}) and Eq.(\ref{part33}) separately  and sum them.  Figs. \ref{oscill}a,b show that the terms with the even and the odd number of photons in the channels do not interfere at the beam splitter, contrary to the terms with the even number of photons in the channels, which is reflected in  Fig. \ref{oscill}c.

\section{$P_{22}$ probability: normalization strategy}
In this section we present the calculation performed to normalize the experimental $P_{22}$ probability. In the experimental setup we used four detectors, namely two detector pairs for every output of the beam splitter. The detectors in each pair are separated by a further beam splitter in two sub-channels (see Fig.\ref{exsetup}). Indicating the detectors as A, B, C, D (A and B correspond to the first channel, C and D correspond to the second channel), we counted all possible combinations. In that sense, $c_{ABC}$ is the number of counts due to the clicks of detectors A, B and C, whereas $c_{ABCD}$ represents the number of counts when all detectors click at the same time. 


In order to estimate photon losses, we add a beam splitter with transmission parameter $\eta$  in each of four output ports before detectors in the detection stage.  The transmission parameter $\eta$ of these beam spitters hence represents the probability for the single photon to reach a detector.

At first step, we estimate the number of photon quadruplets generated in the PDC section $N_4$  and $\eta$. From theory, we calculate that for infinite time delay the coincidence probabilities are given by $P^{th}_{22}=0.37$ and $P^{th}_{31}=P^{th}_{13}=0.25$, so we write down the equations for $c_{ABC}$ and $c_{ABCD}$:
\begin{align}
\langle c_{ABCD}(\tau \gg 0)\rangle=\frac{1}{4} \eta^4 N_4 P_{22}^{th}; \\
\langle c_{ABC}(\tau \gg 0)\rangle= N_4 \bigg [\eta^4 \bigg(\frac{3}{8}P^{th}_{13}+\frac{1}{8}P_{22}^{th}\bigg)+\eta^3(1-\eta)\bigg(\frac{1}{4}P_{13}^{th}+\frac{1}{4}P_{22}^{th}\bigg)\bigg].
\end{align}
By solving simultaneously these two equations we get the experimental values of $N_4$ and $\eta$ which are the same for both the infinite time delay zone and the zero-delay zone, since we used the same detectors and the counts are normalized respect to the pump power and the integration time. Now we can estimate the experimental probability $P^{ex}_{22}$ in one of the peaks of the zero-delay zone by using the proper number of counts:
\begin{equation}
\langle c_{ABCD}(\tau\simeq 0)\rangle=\frac{1}{4} \eta^4 N_4 P_{22}^{ex};
\end{equation}
This strategy, repeated over the four possible detector combinations, provides the final value $P_{22}^{max}\simeq0.915$.
\end{document}